\documentclass[article,prl,balancelastpage,twocolumn,showpacs]{revtex4}
\usepackage{makeidx}
\usepackage{amssymb}
\usepackage{dcolumn}
\usepackage{graphicx}
\usepackage{acronym}

\input{tcilatex}

\begin{document}

\title{Optical measurement of a fundamental constant with the dimension of
time}
\author{Boris V. Gisin }
\affiliation{IPO, Ha-Tannaim St. 9, Tel-Aviv 69209, Israel. E-mail:
borisg2011@bezeqint.net}
\date{\today }

\begin{abstract}
We consider the concept of a rotating reference frame with the axis of
rotation at each point and the applicability of this concept to different
areas of physics. The transformation for the transition from the resting to
rotating frame is assumed to be non-Galilean. This transformation must
contain a constant with dimension of time. We analyze different
possibilities of experimental testing this constant in optics, as most
suitable field for measurements presently, and also in general relativity
and quantum mechanics.
\end{abstract}

\pacs{04.20.Cv, \ 06.20.Jr, \ 42.15.Eq \hspace{20cm}}
\maketitle

\section{Introduction}

The concept of a "point rotation reference frame", i.e., the frame with the
axis of rotation at every point, arises in optics. However, this concept is
also applicable to other areas of physics. An example of such a frame is the
optical indicatrix (index ellipsoid) \cite{Born}.\ Any rotating field,
including spinor and gravitational, is the object of the point rotation.

Coordinates of the frame are the angle, time and axis of rotation. The
radial coordinate is not used in manipulations with the frames. Centrifugal
forces don't exist in such frames. Optically, the rotating half-wave plate
is an equivalent to the resting electrooptical crystal with the rotating
indicatrix \cite{Bur} but, physically, they are different because the plate
has only one axis of rotation. The frames are not compatible with
Cartesian's frames.

The main question in such a concept: what is the transformation for point
rotating reference frames? Is the transformation Galilean or not?

From the viewpoint of contemporary physics a non-Galilean transformation,
with different time for frames rotating at different frequencies, is much
more preferable in comparison with the Galilean, where time is the same.
Moreover, such a transformation must contain a constant with the dimension
of time similarly to the Lorentz transformation and the speed of light. This
constant should define limits of applicability of basic physical laws.\ 

In contrast to mechanics, where the relativity principle is used to deduce
the transformation for the rectilinear motion, such a general principle does
not exist for the point rotation. Therefore, this transformation cannot be
explicitly determined.

It is known that an electric field, rotating perpendicular to the optical
axis of a 3-fold electrooptic crystal, causes rotation of the optical
indicatrix at a frequency equal to half the frequency of the field.\ It
means that the optical indicatrix of such a crystal possesses some
properties of two-component spinor.\ 

The sense of rotation of the circularly polarized optical wave propagating
through this crystal, is reversed, and the frequency is shifted, if the
amplitude of the applied electric field is equal to the half-wave value. The
device for such a shifting is the electrooptical single-sideband modulator 
\cite{pat}.

The use of the transformation makes the description of the light propagation
in the electrooptical single-sideband modulator simpler and comprehensible.

For the description of the phenomenon transit to a rotating reference frame
associated with axes of the indicatrix. As result of such a transition, the
frequency of the wave is shifted by half frequency of the modulating
electric field. This shift is doubled at the modulator output due to the
polarization reversal and transition to the initial reference frame \cite%
{pat} .

In this paper we study the general form of two-dimensional non-Galilean
transformation and the possibility of its experimental verification.
Emphasize, experiment always involves the direct and reverse transformation
because an observer rotating at each point does not exist.

\section{The transformation}

The general form of \ the normalized non-Galilean transformation may be
written as follows%
\begin{equation}
\tilde{\varphi}=\varphi -\Omega t,\text{ \ }\tilde{t}=-\tau \varphi +t,
\label{trn1}
\end{equation}%
where the tilde corresponds to the rotating frame, $\tilde{\varphi},\varphi $
and $\tilde{t},t$ are the normalized angle and time, $\Omega $ is the
frequency of the rotating frame (the modulating frequency is $2\Omega $), $%
\tau (\Omega )$ is a parameter with the dimension of time. The reverse
transformation follows\emph{\ }from (\ref{trn1}) 
\begin{equation}
(1-\Omega \tau )\varphi =\tilde{\varphi}+\Omega \tilde{t},\text{ \ }%
(1-\Omega \tau )t=\tau \tilde{\varphi}+\text{\ }\tilde{t}.  \label{tr1}
\end{equation}

Consider a plane circularly polarized light wave propagating through the
modulator. Transit into the rotating frame. The optical frequency in this
frame is%
\begin{equation}
\tilde{\omega}=\frac{\omega -\Omega }{-\tau \omega +1},  \label{f1}
\end{equation}%
where the frequency in the resting and rotating frame is defined as $\omega
=\varphi /t$ and $\tilde{\omega}=\tilde{\varphi}/\tilde{t}$ respectively.

If the half-wave condition is fulfilled, the reversal of rotation occurs at
the modulator output. For the circularly polarized wave the negative sign of
the frequency corresponds to the opposite rotation. Making transition into
the resting frame with changing the sign of $\tilde{\omega}$, obtain the
output frequency as a function of $\omega $ and $\Omega $ 
\begin{equation}
\omega ^{\prime }=\frac{-\omega (1+\tau \Omega )+2\Omega }{-2\tau \omega
+\tau \Omega +1}.  \label{fr1}
\end{equation}

In fact, we consider the single-sideband modulator in this approach as a
black box. This box changes the sense of rotation of the circularly
polarized light wave and shifts its frequency.

\section{Optics}

For the evaluation of the parameter $\tau $ we use results of optical
measurements from the work \cite{jpc}. In this work the principle of the
single-sideband modulation was checked and the Galilean transformation was
used for the theoretical description of the process.

Circularly polarized light from Helium-Neon laser was modulated by a Lithium
Niobate single-sideband modulator at the frequency 110 MHz. The experiment
showed an asymmetry of the frequency shift for two opposite polarizations.
The extra shifts was of the order of few MHz.

Proof. W. H. Steier, one of the authors of the work \cite{jpc}, kindly
answered my question about the origin of this asymmetry: "Your are correct
about the apparent asymmetry. We never noticed it earlier. I do not know if
this is a property of the scanning mirror interferometer. It has been many
many years since we did that work and all of the equipment has now been
replaced. It would not be possible for us to redo any work or start the
experiments again".

Possibly, the origin of this extra shift is a defect of the equipment. In
any case this shift can be used for approximate estimates of the upper
boundary of the parameter $\tau $. From this the important conclusion
follows. The parameter $\tau $ is very small.

For small \ $\tau $ and $|\Omega |\ll |\omega |$\ the output frequency (\ref%
{fr1}) may be written as%
\begin{equation}
\omega ^{\prime }\approx -\omega +2\Omega +2\tau \omega ^{2}.  \label{frs}
\end{equation}%
The extra shift equals $2\tau \omega ^{2}.$

The exact form of the dependency $\tau (\Omega )$ is unknown. Therefore,
assume that $\tau $ may be expanded in power series in $\Omega $%
\begin{equation}
\tau =\tau _{0}+\tau _{1}^{2}\Omega +\tau _{2}^{3}\Omega ^{2}+\ldots
\label{tau}
\end{equation}%
In such a form all the coefficients $\tau _{n}$ have the dimension of time.

Since $\tau $ is very small and, usually, $\Omega \ll $ $\omega $, we can
restrict ourselves only the first non-zero term in the expansion (\ref{tau}).

\subsubsection{The case $\protect\tau _{0}\neq 0.$}

This case is most favorable for optical measurements from the viewpoint of
simplicity. The constant $\tau _{0}$\ may be called "quantum of time". The
value of the extra shift defines the upper boundary of the quantum of time $%
\sim 10^{-23}\sec $. That corresponds to a distance of the order of the
proton size.

The experiment, similar to \cite{jpc}, provide an excellent opportunity to
measure the parameter $\tau _{0}$. Accuracy of the measurement can be
increased by several orders of magnitude by using modern technology. The
advent of laser cooling has underpinned the development of cold $Cs$
fountain clocks, which now achieve frequency uncertainties of approximately $%
5\cdot 10^{-16}$ and even lesser \cite{NPL}. That can be used in the
measurement. Best accuracy may be achieved in a ring schematic similarly
measurements of the anomalous magnetic moment of electron.

\subsubsection{The case $\protect\tau _{0}=0$}

If $\tau _{0}$ is exactly equal to zero, the accuracy should be increased by 
$1/(\tau _{1}\Omega )$ times. Accordingly to results of \cite{jpc}, the
upper boundary of $\tau _{1}$ is$\ \sim 10^{-16}\sec $. Using the optical
range for the modulation is connected with the problem of phase matching 
\cite{pat}.

Below briefly summarized results of the analysis and possibilities of
measurements in other areas of physics.

\section{General relativity}

Consider the case $\tau _{0}=0$ in application to general relativity. Now we
restrict ourselves only by the second term of $\tau (\Omega )$ and consider (%
\ref{trn1}) as the Lorentz transformation. Usually this name relates to the
rectilinear motion in mechanics. Here the role of the coordinate and
velocity is played by the angle and frequency respectively.

After a normalization

\begin{equation}
(\tilde{\varphi},\tilde{t})\rightarrow \frac{(\tilde{\varphi},\tilde{t})}{%
\sqrt{1+\tau _{1}\Omega }},\text{ \ }(\varphi ,t)\rightarrow (\varphi ,t)%
\sqrt{1-\tau _{1}\Omega },  \label{norm}
\end{equation}%
obtain%
\begin{equation}
\tilde{\varphi}=\frac{\varphi -\Omega t}{\sqrt{1-\tau _{1}^{2}\Omega ^{2}}},%
\text{ \ }\tilde{t}=\frac{-\tau _{1}\Omega \varphi +t}{\sqrt{1-\tau
_{1}^{2}\Omega ^{2}}}.  \label{rL}
\end{equation}%
Analogously mechanics, $\tau _{1}$ can be regarded as the minimum possible
time interval and $1/\tau _{1}$ as the maximum possible frequency.

The form $\ (\tau _{1}^{2}\varphi ^{2}-t^{2})$ \ is invariant under the
transformation (\ref{rL}). \ 

Despite the fact that the Cartesian reference frames are not compatible with
the point rotation reference frames, there exists a solution of Einstein's
equation invariant under the transformation (\ref{rL}). Consider an exact
solution with cylindrical symmetry \cite{Mar}%
\begin{equation}
ds^{2}=Ar^{a+b}dr^{2}+r^{2}d\varphi ^{2}+r^{b}dz^{2}+Cr^{a}dt^{2},
\label{mM}
\end{equation}%
where $A,C,a,b$ are constants. This solutions is a invariant under the
transformation (\ref{rL}) provided $a=2$. Moreover, at $a=b=2$ and a
normalization of $r$ and $t$ the metric can be reduced to the form 
\begin{equation}
ds^{2}=(1+\frac{1}{L}r)[dr^{2}+l^{2}d\varphi ^{2}+dz^{2}-c^{2}dt^{2}],
\label{mp}
\end{equation}%
where\emph{\ }$l\equiv c\tau _{1}$ and $L$ are constants with the dimension
of length. For a "center", at $r=0$, this metric looks like "Euclidean
metric" for the point rotations.

Non-stationary solutions of Einstein's equation, invariant under the
transformation (\ref{rL}), also exists.

The existence of such metrics opens the way for applying the concept of the
point rotation reference frames to general relativity. In this sence
suitable solutions of Einstein's equation are possible but searching for
consequences of such solutions applicable for measurements of $\tau _{1}$ or 
$l$ is not simple problem.

\section{Quantum mechanics}

Initially, quantum mechanics was considered as the most suitable area of
physics for the measurement of the parameter $\tau $. However, the hope to
find in quantum mechanics a consequence of the transformation (\ref{trn1})
applicable to measurements, proved to be illusory.

Quantum states in rotating magnetic or electromagnetic fields are not
stationary. The problem becomes stationary by the transition to the rotating
frame. Main role in the transition plays the phase transformation of the
spinor, which is defined by the first equation in (\ref{trn1}). The second
equation, containing the parameter $\tau ,$ plays a minor role.

The transition was used for finding a new class of exact localized solutions
of the Dirac equation \cite{arx} in the rotating electromagnetic field.

However, the further study showed that the parameter $\tau $ vanishes from
final results due to the reverse transition into the resting frame. It
allows also to conclude that the non-Galilean transformation is not related
to the problem of anomalous magnetic moment\ in any case for the above exact
localized solutions.

\section{Conclusion}

We have considered the concept of the point rotating frame and the
non-Galilean transformation for such frames. The concept is applicable to
optics, general relativity and quantum mechanics. The parameter $\tau $ with
the dimension of time is a distinguishing feature of the non-Galilean
transformation. This parameter is very small.

Presently optics can be considered as the main area of physics for
measurements of the parameter $\tau $.

The nonzero term $\tau _{0}$\ in the expansion (\ref{tau}) is the most
favorable case for optical measurements. However, in the case $\tau _{0}=0$\
measurements are also possible. The experiment would be similar to \cite{jpc}%
, but on the basis of modern technology. Best accuracy may be achieved in a
ring schematic similarly measurements of the anomalous magnetic moment of
electron. This schematic, regardless of results of experiments (positive or
negative), can also be used for high-precise manipulations with the laser
frequency in variety applications, in particular, for standards of length
and time.

A fundamental constant with dimension time must be on the list of basic
physical constants. However, this constant is absent in this list. The
parameter $\tau $\ contains this constant and it should be the basic
physical constant because it is determined by such a basic physical process
as rotation.

The investigation of this problem is very important since the constant\
defines the limits of the applicability of the basic physical laws for very
small intervals of time and length. Moreover, this constant might determine
the minimum possible values of such intervals as well as the minimum
possible value of energy.

The above opinion of prof. W. H. Steier about the origin of the asymmetry is
an argument against funding the high-precise measurements. Nevertheless, the
problem of "to be or not to be" (in the sense of Galilean or not) must be
solved.

\end{document}